# Influence of scanning plane on Human Spinal Cord functional Magnetic Resonance echo planar imaging

Marta Moraschi[1,2], Silvia Tommasin[1,3], Laura Maugeri[1,4], Mauro DiNuzzo[1,5], Julien Cohen-Adad[6], Marco Masullo[1,4], Fabio Mangini [1,7], Lorenzo Giovannelli[1,4], Daniele Mascali[1,5], Tommaso Gili[8], Valerio Pisani[1], Ugo Nocentini[1,9], Federico Giove[1,5,a], Michela Fratini[1,4]

1. IRCCS Fondazione Santa Lucia, Rome, Italy,

2. Operative Research Unit of Radiation Oncology, Campus Bio-Medico University of Rome, Italy

3. Human Neuroscience Department, Sapienza University of Rome, Rome, Italy

4. CNR Nanotec, Rome Italy

5. MARBILab – Museo Storico della Fisica e Centro Studi e Ricerche Enrico Fermi, Rome, Italy

6. NeuroPoly Lab, Institute of Biomedical Engineering, Polytechnique Montreal, Montreal, QC, Canada

7. Department of Information Engineering, University of Brescia, Brescia, Italy

8. Networks Unit, IMT School for Advanced Studies Lucca, Lucca, Italy

9. Clinical Sciences and Translational Medicine Department, University of Rome "Tor Vergata", Rome, Italy

[a]**Corresponding author** : Dr Federico Giove

Magnetic Resonance for Brain Investigation Laboratory, Museo Storico della Fisica e Centro di Studi e Ricerche Enrico Fermi, Rome, Italy.
phone: +39 065150.1324
**e-mail:** federico.giove@cref.it

**ACKNOWLEDGEMENTS**

The authors are grateful to Dr. Claudia Marzi and Dr Marco Clemenzi,  Santa Lucia Foundation, for technical supplying.

**GRANT SUPPORT**

The Italian Ministry of Health Young Researcher Grant 2013 (GR-2013-02358177)

Regione Puglia and CNR for Tecnopolo per la Medicina di Precisione. D.G.R. n. 2117 of 21.11.2018.

## ABSTRACT

**BACKGROUND:** Functional Magnetic Resonance Imaging (fMRI) is based on the Blood Oxygenation Level Dependent contrast and has been exploited for the indirect study of the neuronal activity within both the brain and the spinal cord. However, the interpretation of spinal cord fMRI (scfMRI) is still controversial and its diffusion is rather limited because of technical limitations. Overcoming these limitations would have a beneficial effect for the assessment and follow-up of spinal injuries and neurodegenerative diseases.

**PURPOSE:** This study was aimed at systematically verify whether sagittal scanning in scfMRI using EPI readout is a viable alternative to the more common axial scanning, and at optimizing a pipeline for EPI-based scfMRI data analysis, based on Spinal Cord Toolbox (SCT).

**METHODS:** Forty-five healthy subjects underwent MRI acquisition in a Philips Achieva 3T MRI scanner. $T_2$*-weighted fMRI data were acquired using a GE-EPI sequence along sagittal and axial planes during an isometric motor task. Differences on benchmarks were assessed via paired two-sample t-test at p=0.05.

**RESULTS:** We investigated the impact of the acquisition strategy by means of various metrics such as *Temporal Signal to Noise Ratio (tSNR)*, *Dice Coefficient to assess geometric distortions, Reproducibility* and *Sensitivity*. tSNR was higher in axial than in sagittal scans, as well as reproducibility within the whole cord mask (t=7.4, p<0.01) and within the GM mask (t=4.2, p<0.01). The other benchmarks, associated with distortion and functional response, showed no difference between images obtained along the axial and sagittal planes.

**CONCLUSIONS:** Quantitative metrics of data quality suggest that axial scanning would be the optimal choice. We conclude that axial acquisition is advantageous specially to mitigate the effects of physiological noise and to minimize inter-subject variance.

**Keywords:** spinal cord; functional magnetic resonance imaging; Blood Oxygenation Level Dependent contrast; benchmarks; acquisition protocol

**Abbreviations:**

scfMRI= spinal cord fMRI

OVS= Outer Volume Suppressio

SC= spinal cord

SCT=Spinal Cord Toolbox

MSF= maximum sustainable voluntary contraction force

DSC=Dice Similarity Coefficient

FDR=False Discovery Rate

# INTRODUCTION

The spinal cord (SC) links the brain to the peripheral nervous system, via ascending and descending pathways, and it is involved in several complex functions [1]. Among different imaging methods, functional Magnetic Resonance Imaging (fMRI) represents the most promising tool for non-invasive investigation of the human spinal cord physiological function [2–6]. Spinal cord fMRI (scfMRI) can be also useful for the characterization of diseases, such as those involving traumatic injury, neuroinflammation or neurodegeneration [7–13], or conditions like pain and analgesia [14,15]. However, in spite of these potentialities, the use of scfMRI is fairly restricted in both research and clinical settings [16], due to challenges in data acquisition/analysis [17–19]. Causes and possible mitigation approaches of these problems have been discussed in depth elsewhere [17,20,21]. Briefly, SC physical dimensions, inhomogeneity of surrounding tissues and motion negatively affect data quality and reproducibility.

Even the choice of the basic scanning sequence is not trivial, because of the involved compromises between contrast sensitivity and data quality. In this study, scfMRI acquisitions were performed via Gradient Echo recalled Echo planar Imaging (GE-EPI) with the application of the saturation bands located externally, anteriorly and posteriorly to the spinal cord. While being the standard approach for brain fMRI, GE-EPI in scfMRI is prone to quality issues, including severe distortion and signal cancellation caused by susceptibility changes associated to motion of surrounding tissue. To minimize this effect, various approaches have been proposed. Recently Kinany et al. compared three imaging protocols including either Outer Volume Suppression (OVS) via saturation bands, or inner FOV imaging for selective excitation of the region of interest [22]. The investigate protocols offered similar overall performances, but with different profiles in terms of advantages and drawbacks, suggesting that the imaging protocol must be tailored to take into account the specific experimental aims and overall design. A different approach to cope with the influence of surrounding tissue is acquiring along the sagittal plane with anisotropic voxels to optimize the sampling matrix according to the natural geometry of the SC. However, sagittal scanning may potentially suffer from additional low-quality issues, because of more marked trough-slice inhomogeneity compared to axial scanning, and requires thin slices, generally 2 mm or below. Indeed, GE-EPI is routinely performed with para-axial orientation. However, scanning the SC in axial orientation requires a large number of slices to cover the region of interest along the rostral-caudal direction, thus partially losing the characteristic temporal advantage of the GE-EPI approach. This problem can be mitigated by recurring to simultaneous multislice imaging (SMS) that allows substantial reduction of repetition time for a given number of slices, at the price of somewhat increased sensitivity to motion [23]. Acquisition of a substantial number of slices in rostrocaudal direction may additionally benefit from dynamic shimming [24].

For either orientation, the sampling matrix must be defined to accommodate the surrounding tissue and to avoid phase wrap-around, unless in-plane selective excitation or effective saturation is used.

Irrespectively of the scanning strategy, the development of fMRI analysis methods and relevant software has been almost exclusively restricted to the brain, thus limiting standardization and diffusion of scfMRI. Recently, the Spinal Cord Toolbox (SCT) has been developed [25], and can be used to analyse structural and functional data. SCT provides a standard and common analysis platform suitable also for advanced processing tasks, e.g., by providing the possibility to perform voxel-based group analysis in the PAM50 standard space [26].

This study was aimed at systematically exploring the respective merits of axial and sagittal scanning in scfMRI using EPI readout in a large cohort of healthy subjects. Additionally, we implemented and optimized a pipeline for scfMRI EPI data analysis based on SCT.

## MATERIALS AND METHODS

### Subjects

Forty-five right-handed healthy subjects (median age 29 years, range 20-54 years) participated in this study. Exclusion criteria were diagnosis of neurological disease and contraindication to MRI examination, including psychological disorders like claustrophobia. All enrolled subjects gave informed consent for participating in the study, which was approved by the Local Ethics Committee and was conducted in accordance with the Helsinki declaration.

### MRI protocol

Acquisitions were performed on a Philips Achieva 3T MR scanner (Philips Medical Systems, Best, The Netherlands) using a neurovascular coil array. $T_2$*-weighted fMRI data were acquired using a 2D GE-EPI sequence along sagittal and axial planes, with the following parameters: TE/TR = 25/3000 ms, Flip angle=80°, bandwidth (3000 Hz x 64 pixels) about 1kHz. Sagittal scans had FOV = 192 x 144 x 104 $mm^3$ (IS x AP x RL), acquisition matrix = 64 x 94 x 35 and were reconstructed by the scanner to a voxel size of 1.5 x 1.5 x 2 $mm^3$. Axial scans had FOV = 140 x 140 x 143 $mm^3$ (AP x RL x IS), acquisition matrix 92 x 94 x 34 and reconstructed to a voxel size 1.5x1.5x3 $mm^3$. For both acquisition planes 2 saturation bands were applied (anteriorly and posteriorly to the spinal cord) and a SENSE factor of 2 was used. The order of axial and sagittal scans was randomized between subjects and included 110 EPI volumes per run. Anatomical reference images were acquired using 3D $T_1$-weighted gradient echo sequence (TE/TR = 5.89/9.59 ms, flip angle = 9°, FOV = 240 x 240 x 192 $mm^3$, resolution = 0.75 x 0.75 x 1.5 $mm^3$ interpolated). Heartbeat and respiration data were recorded using scanner integrated plethysmograph and respiratory belt during all functional runs. $T_2$-weighted images were acquired as well and were assessed by an experienced radiologist to check for unknown pathological signs but were not directly used in this work. The imaging FOV included the cervical region of the spinal cord, from C1 to T2.

## Stimulation paradigm

Immediately before the fMRI session, subjects underwent an assessment and training phase outside the MR scanner with the stimulation device (Grip force response device HHSC-1x1-GRFC-V2, CURDES, Philadelphia, US). In a first trial, the maximum sustainable voluntary contraction force (MSF) was determined: subjects were asked to press the device up to their maximum sustainable force with the right hand, and to keep the force for 30s. Then, subjects were trained to perform the task.

The task consisted in a block-designed isometric motor contraction performed with the right hand and guided by a visual feedback system. The stimulation protocol started and ended with a rest epoch, and overall included 11 epochs, alternating task execution and rest. Epochs lasted 30s each, for a total duration of 5.5 minutes. During task epochs, subjects were requested to steadily apply a given level of force on the grip device, pseudo randomly selected among 20%, 40% or 50% of MSF of the subject. The subjects were previously notified that a different amount of force would be needed to reach the target during each epoch. All subjects were asked to perform 4 fMRI runs of the stimulation protocol, two for each acquisition plane.

The Grip force response device was interfaced to a Windows workstation running Presentation software (v. 20.2), allowing real time measurement of the applied force. The same workstation generated the visual feedback and recorded the sampled force, synchronously with the fMRI acquisition. Visual feedback was a red bar, whose length was proportional to the applied force. Following a countdown of 30s, cues about the beginning and end of each epoch were visually presented as well. The target force was always set at mid length of the bar (irrespectively of the actual required force) and identified by yellow marks. Visual feedback was shown on an LCD screen and seen by the subject via a mirror.

To minimize task-related motion, participants were mildly contained with the help of cushions and orthopaedic shoulder restraints. Subjects were instructed to relax the muscles not directly involved in the task, to direct their attention to the feedback bar, and not to move in response to the stimuli throughout both the training phase and the actual fMRI experiment. Subject compliance was continuously visually assessed.

## Pre-processing pipeline

Data pre-processing and processing were performed combining existing software tools commonly employed for brain fMRI with SCT, including slight adaptations of the latter. In particular, we used a custom-made Matlab (The MathWorks, Inc.) implementation of RETROICOR, SPM12 (http://www.fil.ion.ucl.ac.uk/spm/), Analysis of Functional NeuroImages (AFNI) [27], FMRIB Software Library (FSL- https://fsl.fmrib.ox.ac.uk/fsl/fslwiki) and SCT [28] (see figure 1). SCT was used to perform all the spatial preprocessing tasks, while the other software was used for physiological noise mitigation, slice timing, statistical inference and pipeline integration.

Preliminarily, $T_1$-weighted images of each subject were segmented in white matter (WM), grey matter (GM) and cerebrospinal fluid (CSF) using SCT, allowing automated vertebral level identification. A SC mask was obtained merging GM and WM masks. The mean functional image of each run was computed to be used as a reference for realignment after physiological noise correction.

RETROICOR noise correction was applied to each functional run, using recorded pulse and respiration data [29] and including two cardiac and two respiration terms and the relevant first derivatives. Motion correction was then performed with SCT estimating slice-by-slice translations while ensuring regularization constraints along the spine. To avoid metric of the cost function to be weighted towards surrounding structures, motion correction was limited to a region enclosing the spinal cord (about 50 mm x 50 mm in plane) derived from the mean functional image.

Realigned data were corrected for slice-timing using SPM. EPI images were then normalized to the PAM50 template with SCT. The transformation matched fMRI data to the template by automatically generating and registering to the centreline appropriate labels along the SC. Then a multi-step nonlinear warping on the axial-plane was applied to the functional images that were finally registered in rostro-caudal direction using a regularized polynomial function [28].

After normalization, images in PAM50 space were smoothed with a 1D Gaussian Kernel (3 mm) along the spinal cord centreline direction, to cope with natural spinal cord anatomy and minimize CSF contamination.

## Data Analysis

### Functional response

Analysis of fMRI time series was carried out voxel-wise in the time domain by means of a general linear model of the functional response, built by convolving the spinal cord haemodynamic response [30] with the task paradigm.

### Quality Metrics

We investigated the impact of acquisition strategy by means of various metrics.

*Temporal Signal to Noise Ratio (tSNR).* For each subject, the tSNR was evaluated voxel by voxel as the ratio between the mean value and the standard deviation across time of the fMRI series [31]. Mean tSNR across subjects was then averaged within the spinal cord at each vertebral level (from C2 to T2).

*Dice Similarity Coefficient (DSC).* Degree of geometrical fidelity (i.e., a reciprocal measure of distortion) was evaluated for each subject by calculating the DSC [32] between the spinal cord mask derived from the anatomical image (used as undistorted reference) and spinal cord masks derived with SCT from axial or sagittal functional images. DSC provided a measure of the spatial overlap between the two masks.

*Reproducibility* of results across subjects was computed as the Pearson's correlation coefficients between subjects' unthresholded functional t maps within either the GM mask, or the whole cord, thus including both GM and surrounding WM. Significant difference in reproducibility between images was assessed on Fisher-transformed correlation coefficients.

*Sensitivity* was evaluated voxel-wise as the amplitude of functional contrast (BOLD relative change over the baseline) within the GM.

To assess *False Discovery Rate* (FDR) we assumed that functional response should be confined to GM. Activation masks were derived from second level group analysis (two-tail, one sample t-test) in function of different thresholds, and voxel count in GM and WM was weighted with the relevant probabilistic tissue maps.

The number of false positives in GM was estimated by weighting the number of positive voxels in WM with the relative volume of GM/WM. FDR was finally assessed by computing the ratio between the number of false positives and the number of positives in GM.

**Statistics**

Differences on benchmarks were assessed via paired two-sample t-test, significance $p<0.05$.

# RESULTS

Maps of the mean tSNR, calculated across subjects and for both axial and sagittal acquisition planes, are reported in Figure 2a. In general, the average tSNR is greater for images acquired along the axial than the sagittal plane. Moreover, tSNR inside the cord tends to be greater than tSNR within the CSF for the axial plane, whereas tSNR values inside and outside the cord are more comparable in sagittal scans. Analysis in relation to vertebral level showed that tSNR of images acquired along both planes was comparable in high cervical regions (C1-C3), while axial acquisition plane resulted in significantly higher tSNR in lower cervical to upper thoracic levels, thus at level 4 ($t=2.4$, $p<0.02$), level 5 ($t=6.0$, $p<0.001$), level 6 ($t=5.6$, $p<0.001$), level 7 ($t=5.4$, $p<0.001$), level 8 ($t=3.6$, $p<0.001$), level 9 ($t=2.0$, $p<0.05$, Figure 2b).

The degree of overlap between spinal cord masks obtained from T1-weighted and EPI scans is shown for a representative subject in Figure 3. Qualitatively, the higher levels are less affected by distortion, that impacts especially the lower cervical to upper thoracic levels. The spatial fidelity was comparable between strategies: average DSC was 0.43±0.08 for acquisition on the axial plane and 0.44±0.06 for acquisition on the sagittal plane. DSC difference was not significant (t = 0.8, p>0.4).

The inter-subject reproducibility matrices (cross-correlation between GM and WM t-maps of each pair of subjects) are shown in the supplementary Figure S1. Acquisition on the axial plane was more reproducible than acquisition along the sagittal plane within the whole cord mask (t=7.4, p<0.01) and within the GM mask (t=4.2 p<0.01). The histograms of correlation coefficients (displayed in Figure 4) show a displacement towards more positive values for axial scanning for both masks. Indeed, the bins for more positive correlations showed systematically higher frequency for axial strategy, while sagittal scanning prevailed in bins for negative correlation (i.e., opposite results between subjects) or small positive correlation. Mean, variance and skewness were computed for both axial ($mean_{ax}$; $var_{ax}$; $sk_{ax}$) and sagittal plane ($mean_{sag}$; $var_{sag}$; $sk_{sag}$) acquisition from the histogram distribution within the whole cord as well as GM mask. Specifically, it was found that mean, variance and skewness were 0.18, 0.047, 0.25 ($mean_{sag}$, $var_{sag}$, $sk_{sag}$) and 0.059, 0.026, 0.36 ($mean_{ax}$, $var_{ax}$, $sk_{ax}$) when computed across the whole cord while 0.071, 0.065, 0.17 ($mean_{sag}$, $var_{sag}$, $sk_{sag}$) and -0.003, 0.053, 0.42 ($mean_{ax}$, $var_{ax}$, $sk_{ax}$) in the case of GM mask. These results are concordant with what we can observe in Figure 4 being the Sagittal scan distributions more asymmetric than the Axial one and the skewness values higher for the Sagittal scan with respect to Axial one. This trend is present in the case of whole cord mask ($sk_{sag}$ =0.36, $sk_{ax}$=0.25) as well as in the case of GM mask ($sk_{sag}$ =0.42, $sk_{ax}$=0.17). In addition, mean values are closer to zero for the Sagittal scan than the Axial one while variance has higher values for axial scan than for sagittal one distribution for both masks.

Maps of the mean sensitivity across subjects are reported in Figure 5a. Functional contrast is comparable in the axial and sagittal plane. Two-sample paired t-tests performed within the cord at each vertebral level from C3 to C7, the activated region for the force grip task [33], showed no differences between positive BOLD contrast between axial and sagittal strategies (p>0.01 at each level). Figure 5b shows slice by slice (in PAM50 space) the average BOLD response of the most responding voxels (top 33%).

Standard approach for spinal cord fMRI

Finally, assuming WM activations as a proxy for the number of false positives in GM, the FDR was inferior for axial slices at the most relevant thresholds (p=0.01 and 0.001, in particular), while tended to equalize at permissive thresholds, indicating that when false positives increase, their number becomes independent of experimental strategy (Figure 6).

## DISCUSSION

Spinal cord fMRI is affected by many quality issues associated with physiology and technical reasons [17,18,20,34,35]. Spinal cord has small cross-sectional area and variable curvature that impose the voxel size to be small, and the SNR is consequently affected. The signal may be contaminated by signals from surrounding tissue because of motion, off-resonance effects, imperfect excitation or saturation [20]. The proximity of vertebral columns and air-filled organs increases field inhomogeneity and thus distortion. Indeed, the spinal cord is difficult to image with high quality even with anatomical modalities [36], because of the inherent difficulty of MRI technology in coping with the peculiarities of the district. For example, high order shimming is not always performed at 3T, but even when performed it cannot fully compensate for localised field variations [18], a problem exacerbated if the field is variable because of motion of the cord itself or of nearby organs, such as lungs, throat and heart, during cardiac and respiratory cycles. Functional MRI, relying on image intensity time series, is naturally more prone to artefacts associated with signal stability. It has been observed that physiological cyclic movements have a span that may be comparable to the voxel size, especially in the cervical region [17]. Lastly, also CSF flow pulses synchronously with heartbeat, representing another confounding effect when the tissue signal is contaminated by signal from CSF [34].

One of the problems that prevents the diffusion of scfMRI techniques is the lack of standardised protocols for acquisition and processing, especially if EPI based readout is used. Indeed, fast spin-echo based imaging has been extensively and systematically characterized, and the relevant merits have been sufficiently clarified, both in healthy subjects and in patients that may represent additional challenges, including implants ([17], and references therein). Similar efforts have not been devoted to EPI scanning, resulting in the extreme variability of experimental approaches. The need of standardized approaches for structural spinal cord imaging, in the interest of repeatability and multicentric efforts, has been recently stressed [37]. Similar efforts are needed for the systematic optimization of EPI-based scfMRI protocols in functional imaging. This work focused on a specific, but important issue: the effect of the imaging plane. We applied an optimised processing protocol to assess across-subject reproducibility and quality metrics of scfMRI acquired along sagittal and axial planes during a typical block-designed motor task.

**Acquisition protocol**

We implemented a GE-EPI approach for this study. In general, Spin Echo methods produce images of higher quality, especially if using fast spin echo and not EPI readout [18], and are currently used in the majority of published studies [17]. In spite of severe quality shortcomings, there are several reasons to try to extend EPI to spinal cord functional studies. GE-EPI has better temporal resolution than fast spin echo, an important feature when disentangling physiological noise from signal [38], whose relevance is expected to increase with the diffusion of functional connectivity studies in the spinal cord [39,40]. Moreover, GE images produce a T2* contrast, and are thus maximally sensitive to dephasing associated to both BOLD effect and magnetic field inhomogeneity. T2-weighting is in principle less sensitive to BOLD effect, especially at 3T, even if it has been proposed that in the specific conditions of the spinal cord the sensitivity is overall matched [17]. Finally, the use of EPI readout would allow to easily exploit the high number of features optimized for functional imaging that are currently implemented only in EPI sequences. Spin echo imaging, resulting in T2-weighting, has been shown to be less contaminated by signal from superficial draining veins [41–44], however in the spinal cord most large vessels are separated from grey matter by white matter, and thus the relevance of this advantage is questionable. Masking GE-EPI images inside the spine provided a good strategy to avoid the BOLD signal to be dominated by spurious signals coming from the spinal cord surface. The advantage of shorter TR in EPI could be dramatically increased by using multislice excitation, however the increased sensitivity to motion of accelerated techniques makes unclear if the advantages of more acquired data and more effective physiological noise treatment overcome the penalization of more motion artefacts.

The selection of an imaging plane is a crucial step for an fMRI study, with potentially deep impact on data quality. Indeed, homogeneity of the magnetic field, tissue anatomy and sources of physiological noise are deeply anisotropic when dealing with the spinal cord [45,46]. Axial and Sagittal slices are the two natural candidates for EPI each coping with some of the anatomic peculiarities of the spinal cord. Coronal slices are less suited to follow the anterior-posterior curvature of the spinal cord and require more slices to accommodate single subject specific features), Surprisingly, this study did not evidence an indisputable advantage of either of the two approaches. This result can be probably associated with the fact that both strategies have advantages and drawbacks. For example, in thoracic areas R-L phase encoding in axial slices allows avoiding the spreading of noise from lungs to the spinal cord, but requires either a larger matrix or an effective suppression of the signal from the body outside FOV, because of the larger extension of the body in R-L than A-P direction. On the other hand, sagittal slices are intrinsically best suited to avoid aliasing, because of the smaller chest dimension along the A-P direction, but noise associated with lungs or throat motion must be dealt with. Similarly, the low curvature of the spinal cord in the R-L direction allows the use of a reduced number of slices in sagittal scanning, but slices must be comparatively thinner than with axial orientation, the latter offering a trough-slice averaging that has an inherent physiological meaning.

**Data processing: quality benchmarks**

Motion correction is a problematic step in the analysis of functional images of spinal cord, because of the challenging size, flexibility, location among organs whose signals interfere in metric computation, and lack of easily recognizable features in the cranio-caudal direction [47]. Ad-hoc pipelines optimised for spinal cord investigation have been developed to address these issues [18,25,48–50]. We combined spinal cord-optimised motion correction with RETROICOR to cope with the most intense sources of signal variance associated with physiological noise.

After denoising, the tSNR spatial distribution was qualitatively different between acquisition planes. Axial orientation was characterised by higher tSNR in GM, especially at lower cervical levels. Sagittal slices showed instead a relatively uniform distribution of tSNR in the spinal canal, or even a prevalence of peripheral areas, likely biased by the higher local intensity (Figure 2a). The fluctuations of CSF signal are likely to be heavily influenced by partial saturation in axial vs sagittal scans, however temporal SNR averaged within the spinal cord shows differential spatial patterns. It was indeed comparable between planes in the upper cervical levels, while axial slices had a significantly higher tSNR from C4 to T1, levels that encompass the areas expected to respond to the stimulation [51]. Notably, in the protocols used for this study, voxel size is smaller in axial than sagittal slices (6.75 mm$^3$ and 9 mm$^3$, respectively), strongly suggesting that the advantage of axial slices is associated with reduced variance of physiological noise, potentially associated to different partial volume with CSF. This observation is further supported by the anatomically relevant disposition of the effect (relevant in lower areas, likely more impacted by throat and lung motion). Moreover, tSNR tends to be higher on the whole slice in axial scans, including muscular tissue surrounding the cord. This feature suggests overall lower sensitivity to motion for the latter strategy.

The degree of overlap, i.e., DSC, between anatomic reference and functional images indicates that functional images are affected by distortion. In comparison with brain studies, the DSC is substantially smaller, e.g., see [52], an expected feature given the larger magnetic field inhomogeneity and different anatomical properties in spinal cord compared to brain fMRI. The geometric fidelity as quantified by DSC was comparable between acquisition planes both in statistical (p>0.4) and numerical terms (0.43±0.08 and 0.44±0.06, axial and sagittal planes, respectively). Qualitative observation suggests that distortion is unevenly distributed in the craniocaudal direction of axial slices, with lower slices more distorted, while distortion appears more homogeneous across spinal cord levels in the sagittal acquisition.

This work was not aimed at identifying the areas responding to the task or at studying the physiology of the response to a multiple level task, but rather at assessing the effects of acquisition plan on the reliability of functional analysis. Cross-correlation between all subjects unthresholded t-maps is expected to be roughly centred on zero, reflecting the number of non-responding voxels, with a tail towards positive correlations associated to reproducible positives. The tail was better delineated in axial images, but the difference was larger in the whole spinal cord cross section compared to grey matter, suggesting the spreading to WM of time-locked oscillations This is probably due to the intrinsic blurring of the images associated with EPI readout and, potentially, with motion associated to task. of the width of the cross-correlations distribution suggests that the pattern of responses show a marked variability between subjects.

The average BOLD amplitude in the most responding voxels (top 33% computed at each level in PAM50 space) is comparable between axial and sagittal acquisition (Figure 5 a,b), in spite of the different tSNR and reproducibility. However, FDR is lower for axial scanning at the most relevant thresholds (Figure 6), while tends to be comparable at permissive thresholds, in agreement with the hypothesis that axial slices are potentially less influenced by physiological correlated noise (or, alternatively, RETROICOR denoising, is more efficient in axial slices). However, the metrics built upon activation analysis depend on the model itself and on the underlying assumptions. In particular, FDR assessment leverages on the assumption that all the activations in WM are false positives. This assumption is questionable according to several previous studies finding activations in WM [39] and to the results of this study, which suggest coherent activity in the WM (see above). The presence of true positives in WM would imply that FDR is overestimated in the present study.

The recent work of Kinany et al indicated that the experimental focus might drive the protocol selection to mitigate efficiently the effects of susceptibility changes in surrounding tissue [37]. While we tested only one of the strategies investigated by Kinany et al, our study highlighted a definite advantage for a specific slice orientation, namely axial scanning. We limited our investigation to standard scanning techniques that are reasonably uniform across vendors, thus we expect that our results are generalizable to other 3T scanners. It should be noted that we tested only a set of parameters for each orientation. Other combinations of geometrical parameters may potentially influence our results, directly affecting SNR, the degree of partial volume contamination, the sampling time, and indirectly influencing important factors like the efficacy of physiological noise suppression and the degree of susceptibility related distortion. However, we don't expect that significantly different geometrical prescription are used, because of the anatomical constraints associated to functional imaging of cervical and thoracic spinal cord. The use of advanced strategies (in particular simultaneous multislice excitation or dynamic z-shimming) is not expected to specifically advantage sagittal acquisition, or it's even impossible in sagittal orientation.

**Limitations.** Many quality-related scanning options can be finely tuned, and extensive exploration of the complete QA-related parameters space is not possible in vivo. Moreover, many strategical choices interact. For instance, sagittal scanning lends itself toward a bigger voxel because of sampling and anatomical constraints, and the largest dimension is bound to be in the slice selection direction for axial slices, and in the readout direction for sagittal slices. Each of these features has consequences on SNR, partial volume, blurring direction. In principle, it is possible to acquire two datasets with identical parameters, except the imaging plane, however these datasets would be suboptimal for at least one, if not both planes, making at least one of the protocols not representative of the actual scanning practice and thus rendering pointless the comparison. For these reasons, it is difficult to generalize the results of this work beyond the set of parameters here explored. However, this work clearly points out that the imaging plane has consequences on data quality, that affects not only the low-level features of the images, but also the features of the functional contrast, including reproducibility and sensitivity.

**Conclusions**

We applied preprocessing tailored to EPI-based fMRI of the spinal cord to systematically investigate the optimal scanning plane. Objective quality metrics indicated a definitive advantage for the acquisition in axial orientation. Reproducibility between subjects, a critical feature in scfMRI, is influenced by the imaging plane and is better in axial images with our protocols, suggesting that axial slices may be the preferred choice in studies on healthy humans as well as in clinical studies.

**Figures**

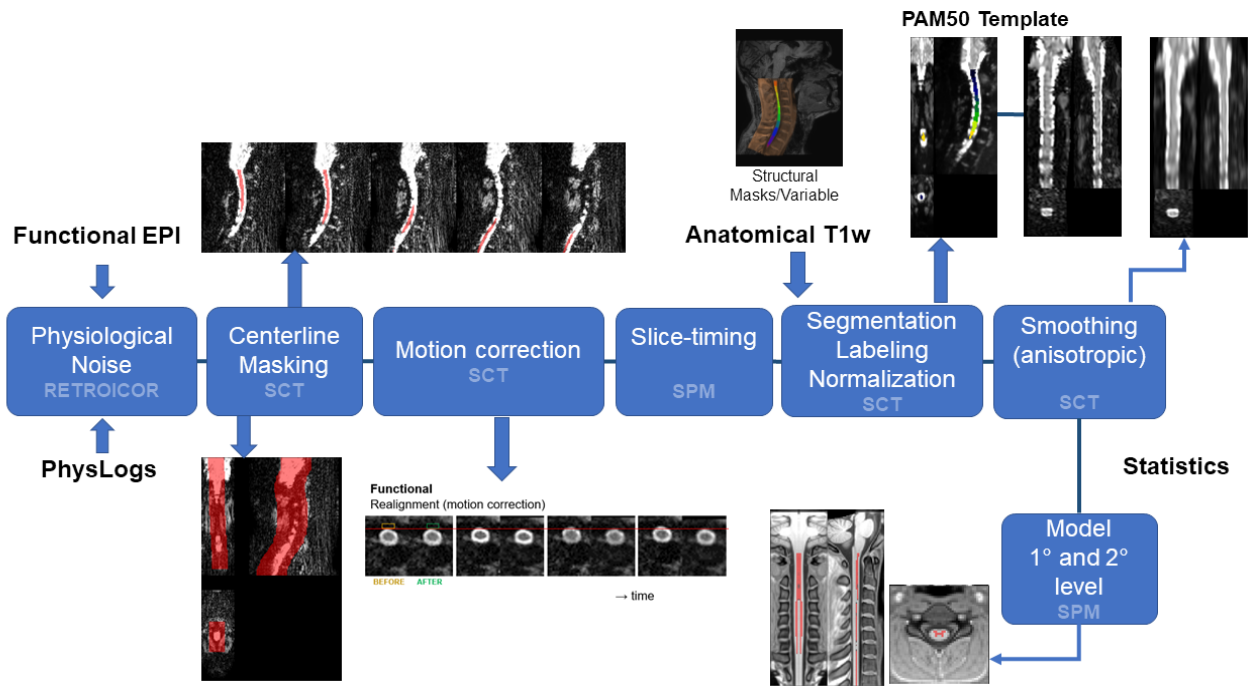

**Figure 1.** Schematics of the data processing pipeline illustrating the main steps and related toolbox used (SCT or SPM). PhysLogs: physiological log files.

Standard approach for spinal cord fMRI

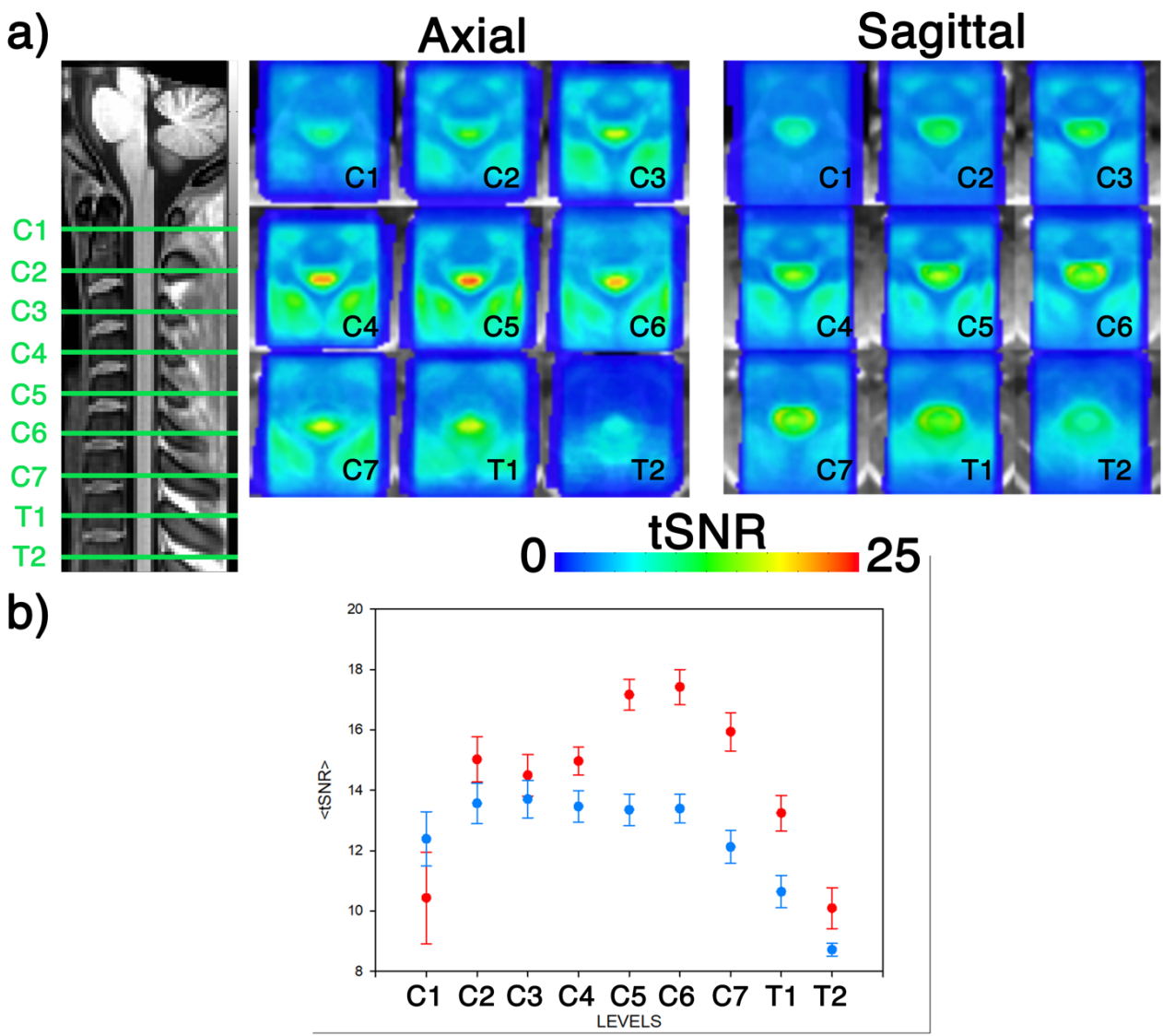

**Figure 2.** a) Maps of the tSNR averaged across subjects in the PAM50 space, at levels from C1 to T2. The maps are overlaid on a standard anatomical template. b) tSNR averaged within the spinal cord at vertebral levels from C1 to T2 for images acquired on axial (red) or sagittal (blue) planes. Data are mean and standard error of the mean across subjects. Significant differences are calculated via two-sample paired t-test and shown as * if p ≤ 0.05, and ** if p < 0.001.

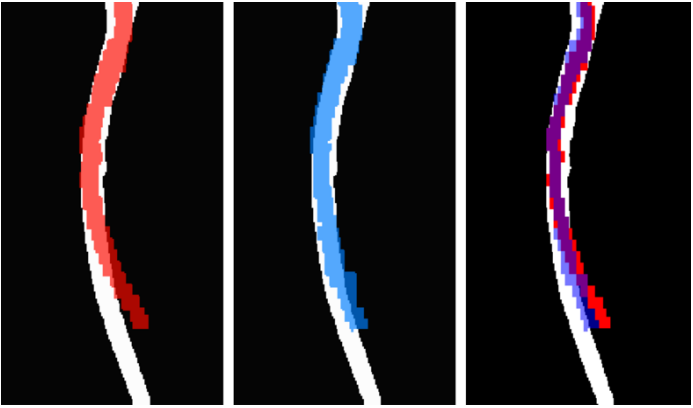

**Figure 3:** Masks of the spinal cord of a representative subject for axial (red, left) and sagittal planes (blue, center). Overlap (right) is highlighted in purple. The relevant T1 mask, assumed as undistorted reference, is white. Average DSC was 0.43±0.08 for acquisition on the axial plane and 0.44±0.06 for acquisition on the sagittal plane. DSC difference was not significant (t = 0.8, p>0.4) between scan planes. Beyond distortion, EPI masks have a generally smaller cross-sectional area because EPI masking was performed aggressively to minimize contamination by CSF.

# REPRODUCIBILITY

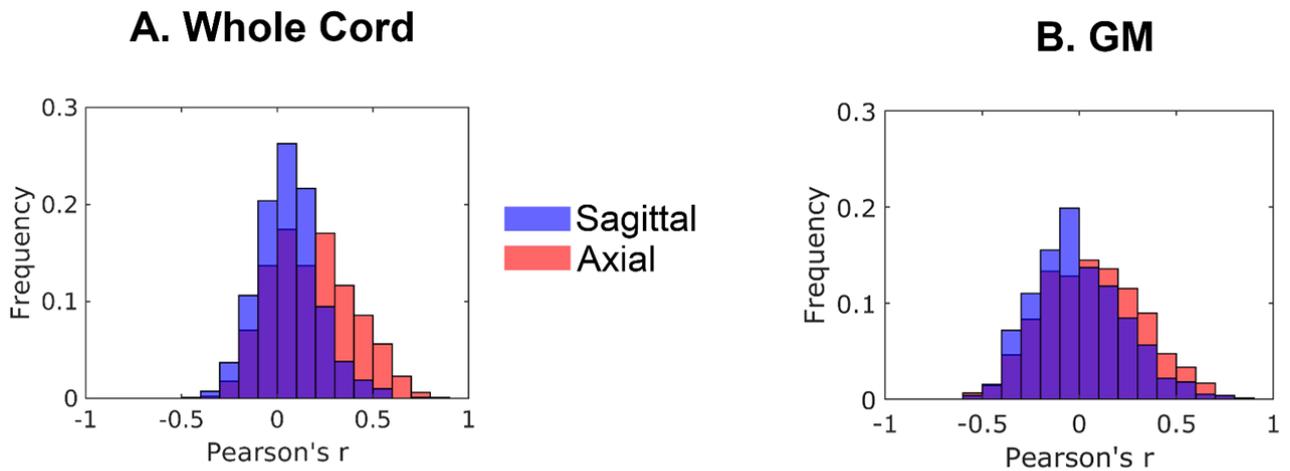

**Figure 4:** Reproducibility assessed as inter-subject correlation matrices of unthresholded t maps in spinal cord cord (A), including gray matter (GM) and surrounding white matter, and in central gray matter (B). From the inter-subject correlation matrices the frequency of Pearson's correlation coefficient values was calculated and displayed in the figure (red: axial, blue: sagittal).

## A. BOLD response maps

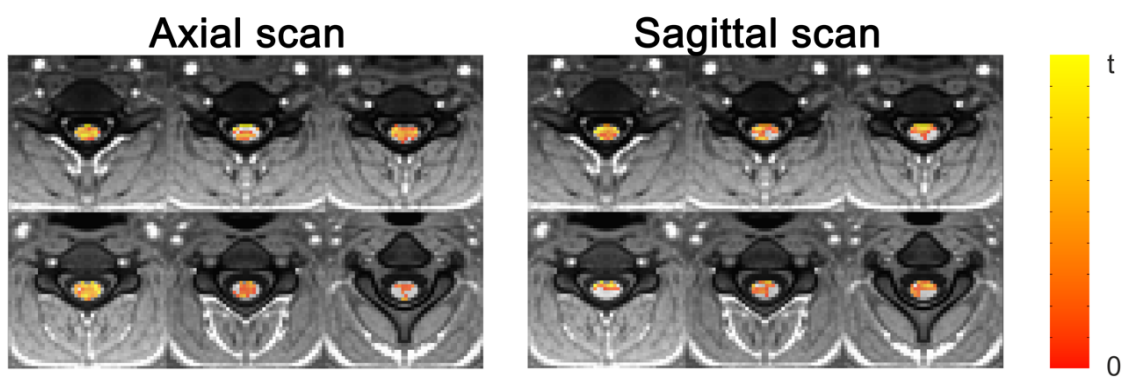

## B. Average BOLD response

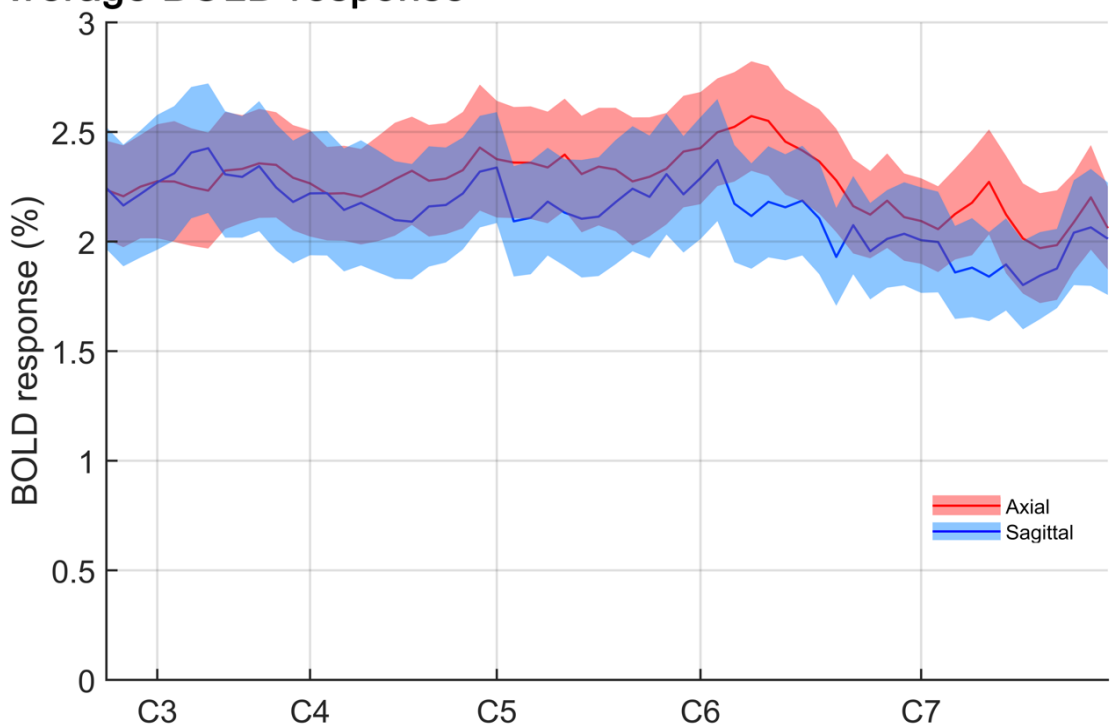

**Figure 5:** A) Maps of the positive functional contrast (BOLD response) at levels C3-C7 in axial (left) and sagittal (right) acquisition (t=2.4, p>0.01). B) BOLD response (top 33% most responding voxels) averaged at different rostrocaudal levels, in PAM50 space. The response is expressed as percent of the baseline. The approximate position of the center of each spinal segment is marked on the horizontal axis. The amplitude of BOLD in most responding voxels in axial and sagittal acquisition is mostly overlapped at each spinal level.

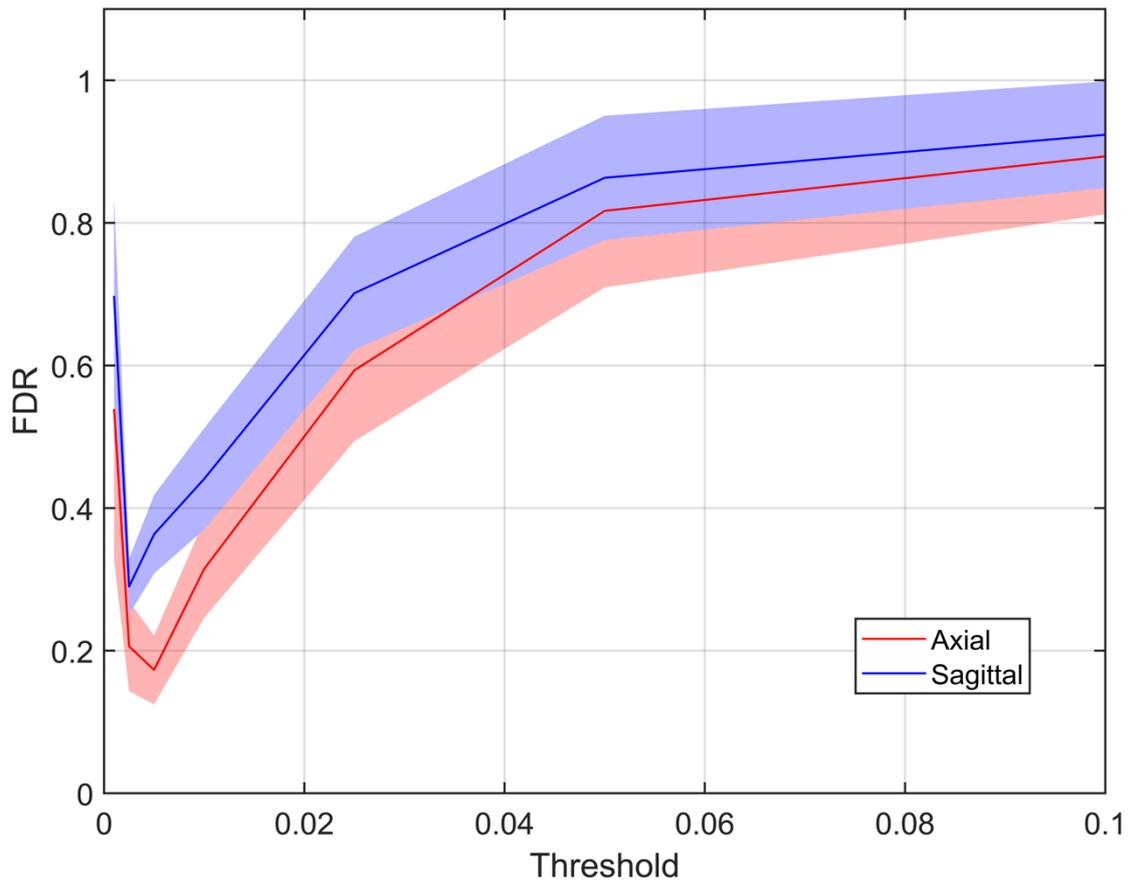

**Figure 6:** False Discovery Rate was estimated assuming that the fraction of active voxels in WM is repetitive of the fraction of False Positives in gray matter (GM) and displayed as a function of threshold p-value, for images acquired on axial (red) and sagittal (blue) planes.

# REPRODUCIBILITY

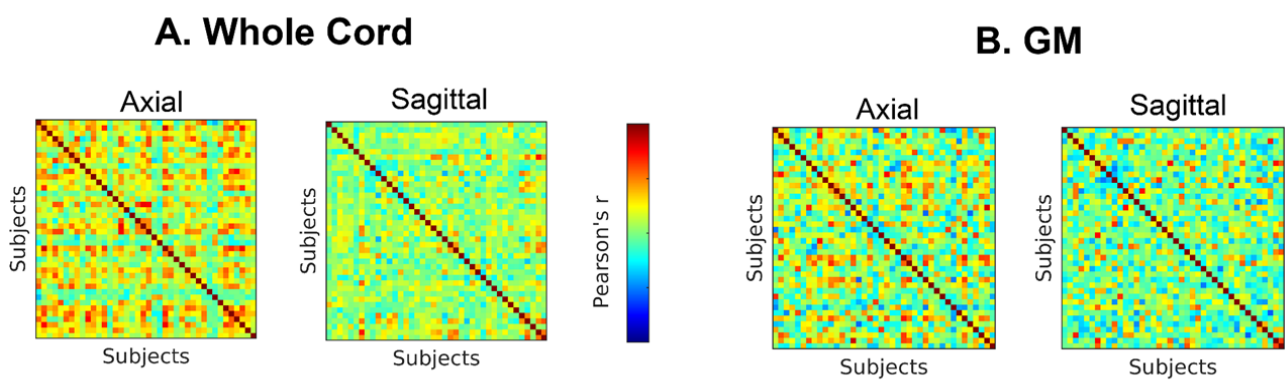

**Supplementary Figure:** Inter-subject correlation matrices of unthresholded t maps in spinal cord cord (A), including gray matter (GM) and surrounding white matter, and in central gray matter (B) are used to assess reproducibility and are shown for axial and sagittal planes.